\documentclass[a4paper]{article}

\usepackage{INTERSPEECH2021, tablefootnote, stfloats, multirow}

\def\s{{\mathbf s}}
\def\c{{\mathbf c}}
\def\z{{\mathbf z}}
\def\x{{\mathbf x}}
\def\L{{\cal L}}

\title{UnivNet: A Neural Vocoder with Multi-Resolution Spectrogram Discriminators for High-Fidelity Waveform Generation}
\name{Won Jang$^1$, Dan Lim$^2$, Jaesam Yoon$^1$, Bongwan Kim$^1$, Juntae Kim$^1$}
\address{$^1$Kakao Enterprise Corporation, Seongnam, Republic of Korea\\
$^2$Kakao Corporation, Seongnam, Republic of Korea}
\email{$^1$\{taylor.martin, jeff.rey, montae.k, jaytee.k\}@kakaoenterprise.com\\
$^2$satoshi.2018@kakaocorp.com}

\begin{document}

\maketitle

\begin{abstract}
Most neural vocoders employ band-limited mel-spectrograms to generate waveforms. If full-band spectral features are used as the input, the vocoder can be provided with as much acoustic information as possible. However, in some models employing full-band mel-spectrograms, an over-smoothing problem occurs as part of which non-sharp spectrograms are generated. To address this problem, we propose UnivNet, a neural vocoder that synthesizes high-fidelity waveforms in real time. Inspired by works in the field of voice activity detection, we added a multi-resolution spectrogram discriminator that employs multiple linear spectrogram magnitudes computed using various parameter sets. Using full-band mel-spectrograms as input, we expect to generate high-resolution signals by adding a discriminator that employs spectrograms of multiple resolutions as the input. In an evaluation on a dataset containing information on hundreds of speakers, UnivNet obtained the best objective and subjective results among competing models for both seen and unseen speakers. These results, including the best subjective score for text-to-speech, demonstrate the potential for fast adaptation to new speakers without a need for training from scratch.
\end{abstract}

\section{Introduction}
Vocoders have been employed in various fields such as text-to-speech\cite{shen2018natural, lim2020jdi, ren2020fastspeech, lancucki2021fastpitch}, voice conversion\cite{liu2018wavenet}, and speech-to-speech translation\cite{jia2019direct}. Neural vocoders based on deep neural networks can generate human-like voices, instead of using traditional methods that contain audible artifacts\cite{griffin1984signal, kawahara1999restructuring, morise2016world}. Recently, a few attempts have been made to apply generative adversarial networks (GANs)\cite{goodfellow2014generative} in vocoders to generate more realistic waveforms\cite{kumar2019melgan, yamamoto2020parallel, kong2020hifi, zeng2021lvcnet, mustafa2021stylemelgan}. This new methodology, involving a generator and discriminator, has allowed for waveforms to be generated in the vocoders at a considerably higher real-time speed than in models with autoregressive constraints\cite{van2016wavenet, kalchbrenner2018efficient}, and also, waveforms can be generated with reasonable fidelity and fewer parameters than those required for models based on bijective flow\cite{prenger2019waveglow, ping2020waveflow}.

Most neural vocoder implementations employ band-limited mel-spectrograms to generate waveforms\cite{yamamoto2020parallel, kong2020hifi, zeng2021lvcnet, mustafa2021stylemelgan, prenger2019waveglow}. In this situation, the corresponding acoustic information from the high-frequency band is not provided to the model. If spectral features of up to half of the sampling rate\cite{shannon1949communication} are used as the input, full-band acoustic information can be provided to the model for the reconstruction of high-fidelity waveforms. However, in some models employing full-band mel-spectrograms, an over-smoothing problem occurred, where non-sharp spectrograms have been generated\cite{yamamoto2020parallel}.

Herein, we attempted to solve this problem by using a discriminator. In the field of voice activity detection (VAD), the use of multiple features computed at different spectral and temporal resolutions is helpful for improving binary classification performance under various background noises\cite {zhang2015boosting}. Considering this background, if a discriminator operated as a binary classifier employs multiple resolution spectral features as well as temporal features as the input, we expect that the spectrogram resolution of the generated waveform would improve.

Therefore, we propose UnivNet\footnote{Audio demo samples can be found at the following URL: \\ https://kallavinka8045.github.io/is2021/}, a neural vocoder that synthesizes high-fidelity waveforms in real time. We added a multi-resolution spectrogram discriminator (MRSD) that uses multiple linear spectrogram magnitudes computed using various parameter sets. Using full-band mel-spectrograms as the input, we expect to generate high-resolution signals over the full-band using the MRSD. Our overall discriminators, which combine a multi-period waveform discriminator (MPWD)\cite{kong2020hifi} employing multiple scales of the waveforms, are expected to support the vocoder generating fine-grained waveforms by modeling both spectral and temporal domains.

We devised an evaluation scenario, in which the model was trained from scratch using a large multi-speaker English dataset. UnivNet exhibited the best scores among GAN-based vocoders in all objective and subjective evaluations, regardless of whether the speaker of the input feature is seen or unseen. For text-to-speech evaluation, in which the model was fine-tuned with the predicted features from an acoustic model, UnivNet exhibited the best subjective evaluation score, which demonstrates that it can rapidly adapt to a new speaker without having to be trained from scratch.

\section{Related work}
MelGAN\cite{kumar2019melgan} is one of the first vocoders with which mel-spectrogram inversion using a GAN setup consisting of a non-autoregressive generator and a multi-scale waveform discriminator has been successfully achieved. In the case of Parallel WaveGAN\cite{yamamoto2020parallel}, both fast model convergence and real-time inference were achieved without a distillation process, via training of a non-autoregressive WaveNet\cite{ping2018clarinet} with the weighted sum of the adversarial loss term and the multi-resolution short-time Fourier transform (STFT) loss term. In the case of HiFi-GAN\cite{kong2020hifi}, a multi-period waveform discriminator was proposed that model various periodic patterns of input waveforms; thereby, signals with a quality surpassing that of autoregressive or flow-based models have been generated.


\begin{figure*}[t!]
\begin{minipage}[t!]{0.49\linewidth}
  \centering
  \centerline{\includegraphics[height=7.1cm]{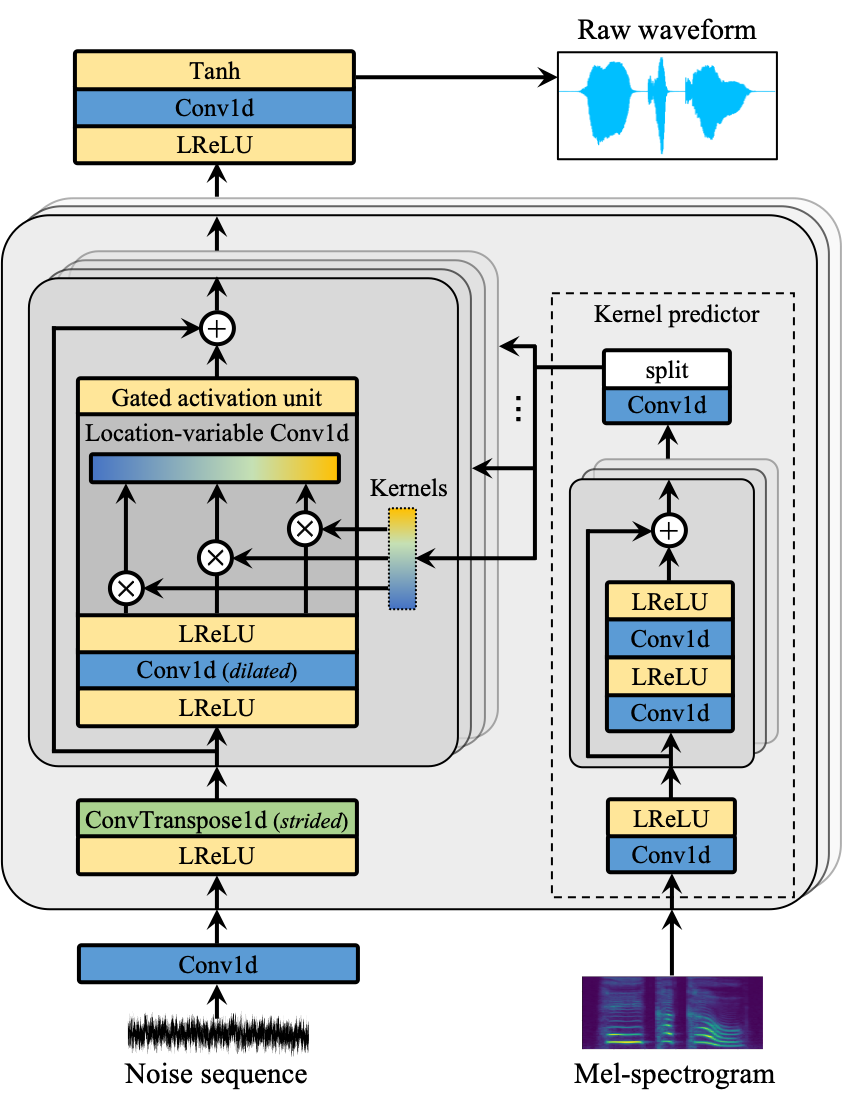}}
  \centerline{(a) Generator}\medskip
\end{minipage}
\begin{minipage}[t!]{0.49\linewidth}
  \centering
  \centerline{\includegraphics[height=7.1cm]{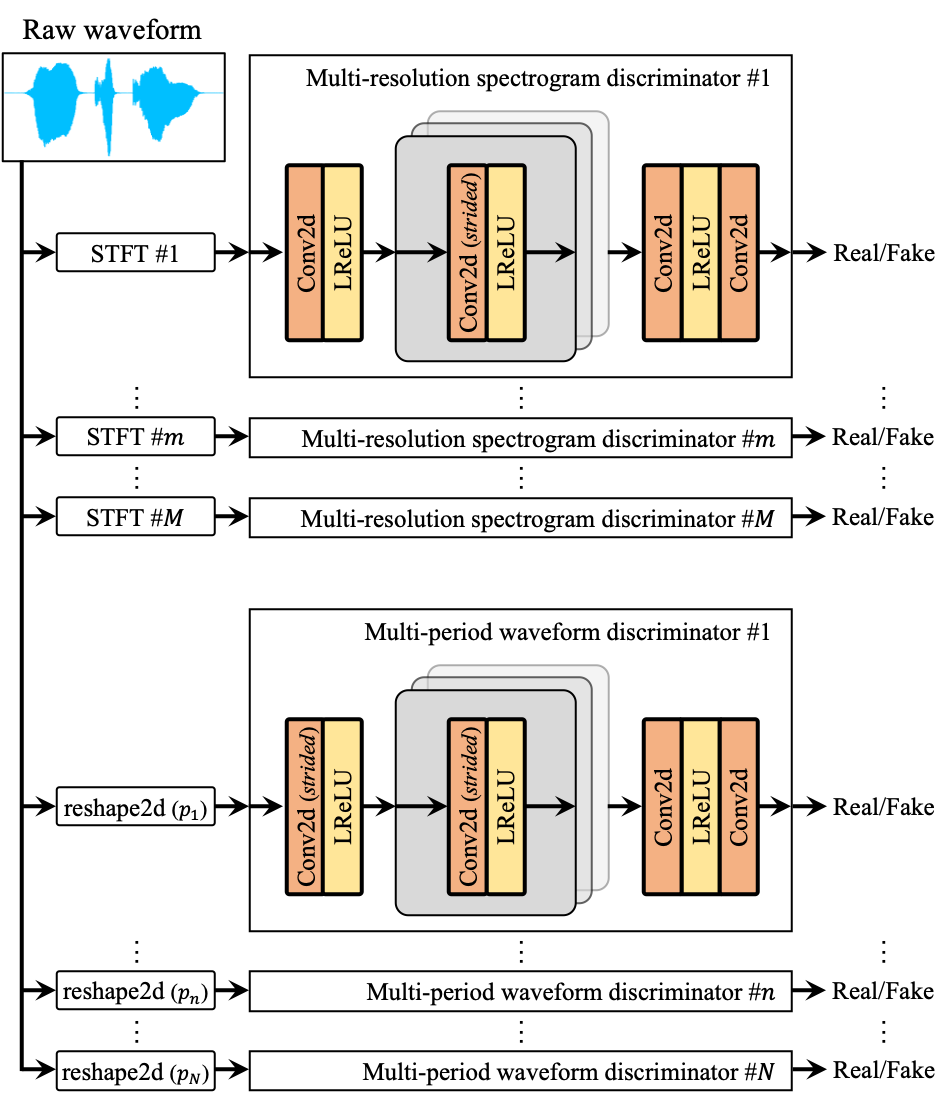}}
  \centerline{(b) Discriminators}\medskip
\end{minipage}
\caption{UnivNet architecture. ‘$\mathrm{STFT\ }\#m$’ denotes the process of computing a linear spectrogram magnitude using the $m$-th STFT parameter set. ‘$\mathrm{reshape2d(}p\mathrm{)}$’ denotes the process of reshaping a 1-D signal of length $T$ to a 2-D signal of height $T/p$ and width $p$.}
\end{figure*}

\section{Description of the proposed model}

\subsection{Generator}
Our generator $G$ is inspired by MelGAN\cite{kumar2019melgan, yang2020multi}. A noise sequence $\z$ is used as the input, and a log-mel-spectrogram $\c$ is used as the condition. The length of $\z$ is the same as that of $\c$, and the length of the output $\hat{\x}$ is the same as that of the target waveform $\x$ owing to transposed convolutions.

To efficiently capture the local information of the condition, location-variable convolution (LVC)\cite{zeng2021lvcnet} was added, which obtained better sound quality and speed while maintaining the model size. The kernels of the LVC layers are predicted using a kernel predictor that takes the log-mel-spectrogram as the input. As shown in Figure 1(a), the kernel predictor is connected to a residual stack. One kernel predictor simultaneously predicts the kernels of all LVC layers in one residual stack. Through internal experiments, we carefully decided on the optimal location and number of the LVC layers as well as the kernel predictors in terms of sound quality and generation speed. To improve the generality of the multi-speaker dataset, a gated activation unit (GAU)\cite{van2016conditional} was added to each residual connection. It was effective in increasing the nonlinearity in WaveNet\cite {van2016wavenet}, and it has also been used in various other studies\cite{yamamoto2020parallel, zeng2021lvcnet, prenger2019waveglow, ping2020waveflow, ping2018clarinet}.

\subsection{Discriminators}
Our discriminators $D$ shown in Figure 1(b) employ multiple spectrograms and reshaped waveforms computed from real or generated signals. First, we propose a multi-resolution spectrogram discriminator (MRSD). To serve as input for each $m$-th sub-discriminator, $M$ real or generated linear spectrogram magnitudes $\{\s_{m}\!\!=\!\!\vert{}FT_{m}(\x)\vert{},\:\hat{\s}_{m}\!\!=\!\!\vert{}FT_{m}(\hat{\x})\vert{}\}_{m=1}^{M}$ were computed from the same waveform using $M$ STFT parameter sets $\{FT_{m}(\cdot)\}_{m=1}^{M}$, with each including the number of points in the Fourier transform, frame shift interval, and window length. There is a conceptual similar work\cite{mustafa2021stylemelgan} that discriminates one spectral information by dividing it into several bands using a filter bank. The difference is that MRSD employes multiple spectrograms with various temporal and spectral resolutions, so we expect to generate high-resolution signals over the full-band. The model architecture is inspired by a multi-scale waveform discriminator (MSWD)\cite{kumar2019melgan} and consists of strided 2-D convolutions and leaky rectified linear unit (LReLU)\cite{maas2013rectifier}.

To improve detailed adversarial modeling in the temporal domain, we added a multi-period waveform discriminator (MPWD)\cite{kong2020hifi}. The periodic components of the waveform are extracted at intervals of a set of prime numbers and used as the input to each sub-discriminator.

\subsection{Training losses}
Multi-resolution STFT loss\cite{yamamoto2020parallel} is used as an auxiliary loss to train the model; it corresponds to the sum of multiple spectrogram losses computed using various STFT parameter sets. The loss $\L_{\mathrm{aux}}$, comprising the spectral convergence loss $\L_{\mathrm{sc}}$ and log STFT magnitude loss $\L_{\mathrm{mag}}$, is defined as follows:
\begin{equation}
\L_{\mathrm{sc}}(\s, \hat{\s}) = \frac{\Vert{}\s-\hat{\s}\Vert{}_{F}}{\Vert{}\s\Vert{}_{F}},\: \L_{\mathrm{mag}}(\s, \hat{\s}) = \frac{1}{S}\left\Vert\log{\s}-\log{\hat{\s}}\right\Vert_{1}
\end{equation}
\begin{equation}
\L_{\mathrm{aux}}(\x,\hat{\x})=\frac{1}{M}\sum_{m=1}^M\mathbb{E}_{\x,\hat{\x}}\Big{[}\L_{\mathrm{sc}}(\s_{m},\hat{\s}_{m})+\L_{\mathrm{mag}}(\s_{m},\hat{\s}_{m})\Big{]}
\end{equation}
where $\Vert\cdot\Vert_{F}$ and $\Vert\cdot\Vert_{1}$ denote the Frobenius and L1 norms, respectively, and $S$ denotes the number of elements in the spectrogram. Each $m$-th $\L_{\mathrm{sc}}$ and $\L_{\mathrm{mag}}$ reuse $\s_{m}$ and $\hat{\s}_{m}$ used in the $m$-th MRSD sub-discriminator. The number of each loss is $M$, which is the same as the number of MRSD sub-discriminators.

We used the objective functions of least-squares GAN\cite{mao2017least} for this study. Our overall objectives are defined as follows:
\begin{equation}
\L_{G}=\lambda\L_{\mathrm{aux}}(\x, G(\z,\c))+\frac{1}{K}\sum_{k=1}^{K}\mathbb{E}_{\z,\c}[(D_{k}(G(\z,\c))-1)^2]
\end{equation}
\begin{equation}
\L_{D}=\frac{1}{K}\sum_{k=1}^{K}(\mathbb{E}_{\x}[(D_{k}(\x)-1)^2]+\mathbb{E}_{\z,\c}[D_{k}(G(\z,\c))^2])
\end{equation}
where $D_{k}$ denotes the $k$-th sub-discriminators of MRSD and MPWD, and $K$ denotes the number of all sub-discriminators. $\lambda$ represents a parameter that balances adversarial and auxiliary loss. The sum of the outputs of all sub-discriminators is divided by $K$ to prevent the balance with auxiliary loss from changing according to the number of sub-discriminators.

\section{Experiments}

\subsection{Data configurations}

We devised an evaluation scenario that employs a large multi-speaker English dataset to train the model from scratch. The LibriTTS\cite{zen2019libritts} dataset is an English multi-speaker audiobook dataset. The ‘train-clean-360’ subset consisting of 192 hours' worth of data, 116k utterances, and 904 speakers was used to train the model and evaluate the speakers used for training (seen speakers). Here, 5\% and 2\% of the seen speakers' utterances were randomly sampled for validation and testing. The ‘test-clean’ subset consisting of 9 hours' worth of data, 5k utterances, and 39 speakers was used to evaluate the speakers not used for training (unseen speakers). For text-to-speech evaluation, we fine-tuned the vocoders with the predicted log-mel-spectrograms to reduce feature mismatch. We prepared the LJSpeech\cite{ito2017lj} dataset containing 24 hours' worth of data and 13k utterances of a single English female speaker.

All waveforms were resampled to a rate of 24 kHz. Using a 1024 point Fourier transform, 256 sample frame shift, and 1024 sample Hann window length, 100-band log-mel-spectrograms (0--12 kHz) were computed; they were then globally normalized using the mean and variance of the entire training set.

\subsection{Evaluation metrics}

We prepared two objective evaluations and one subjective evaluation to measure the sound quality of all the models for the experiments. PESQ\cite{rix2001perceptual} (higher is better) is a metric used for the objective evaluation of speech quality. For wideband PESQ evaluation, each sample was downsampled to 16 kHz. The ‘pypesq\footnote{https://github.com/vBaiCai/python-pesq}’ library was used for the evaluation. In addition, we defined ‘RMSE’ which compares the root-mean square errors of the linear spectrogram magnitude between the original and synthesized sample (lower is better). To show the spectrogram inversion performance of each model, we used the same values as those for the STFT parameters employed to compute the log-mel-spectrograms of the training data.

We prepared a mean opinion score (MOS) subjective evaluation in which raters scored the audio quality from 1 (negative) to 5 (positive) for each sample. Here, 20 individual speakers' utterances were prepared for each assessment and evaluated by 20 subjects in the United States located using a crowd-sourced evaluation via Amazon Mechanical Turk. A total of 400 scores were collected to compute the mean and 95\% confidence intervals for each assessment. All MOS samples were normalized to a loudness of -23 LUFS.

\subsection{Model details}
\begin{figure}[t]
\begin{minipage}[t]{0.45\linewidth}
  \centering
  \centerline{\includegraphics[width=1.0\linewidth]{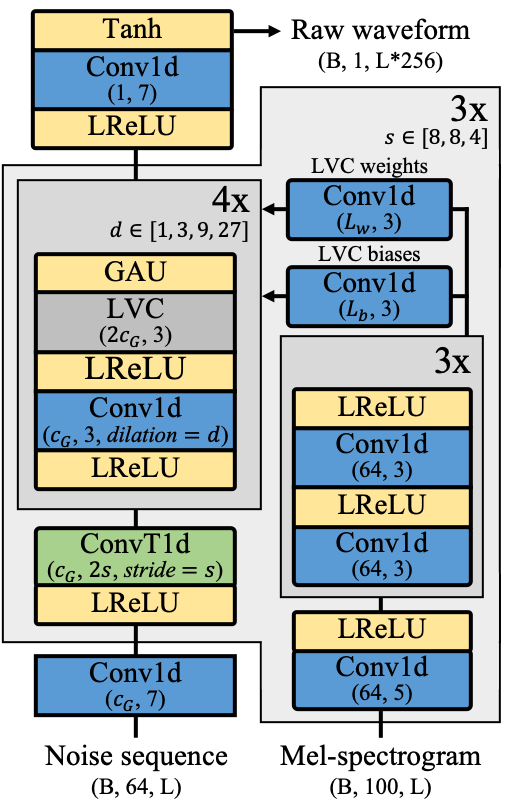}}
  \centerline{(a) Generator}\medskip
\end{minipage}
\begin{minipage}[t]{0.54\linewidth}
  \centering
  \centerline{\includegraphics[width=1.0\linewidth]{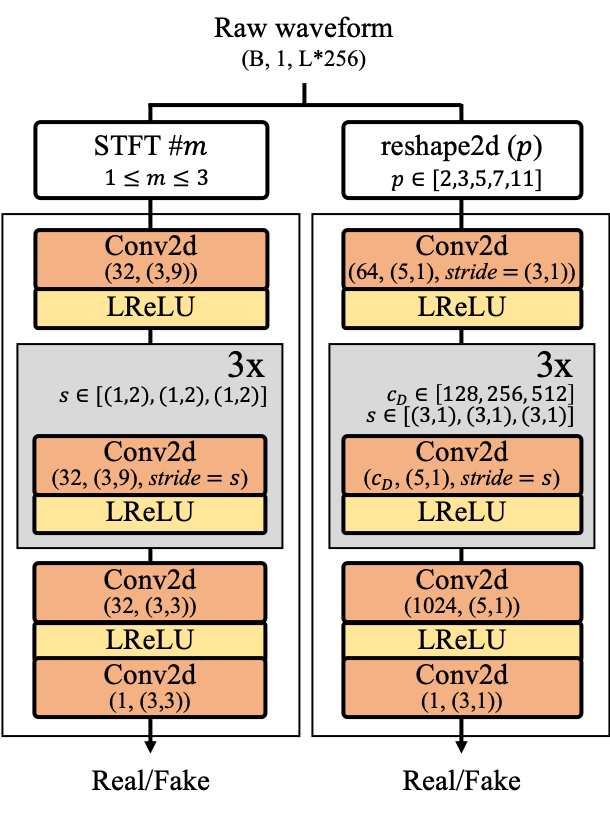}}
  \centerline{(b) Discriminators}\medskip
\end{minipage}
\caption{Details of UnivNet as used in the experiments. The first two numbers in parentheses for each layer indicate the channel size and kernel size, respectively.}
\end{figure}

The details of UnivNet as used in the experiments are shown in Figure 2. We set the dimension of the noise sequence $\z$ to 64. Each element of $\z$ was sampled from a normal Gaussian distribution. All layers in the generator have the same channel size $c_{G}$ except for the last layer, LVC, and the layers in the kernel predictors. We prepared two versions with a channel size $c_{G}$ of 16 and 32, respectively (e.g., UnivNet-c16 and UnivNet-c32). The lengths of the kernel predictor outputs $L_{w}$ and $L_{b}$ are dynamic based on the number of residual connections, input and output channel sizes, and kernel size of LVC. LReLU with $\alpha=0.2$ was used for each activation, and weight normalization\cite{salimans2016weight} was applied to all layers. Three STFT parameter sets (number of points in the Fourier transform, sample length of the frame shift, sample length of the Hann window) for MRSD and auxiliary loss were selected through internal experiments using the validation set, as follows: (1024, 120, 600), (2048, 240, 1200), and (512, 50, 240). The balancing parameter, $\lambda{}$, was 2.5. The model was trained using the Adam\cite{kingma2014adam} optimizer with $\beta_{1}$ = 0.5, $\beta_{2}$ = 0.9, and under a 1$e$-4 learning rate. We trained the generator with only auxiliary loss without discriminators in the first 200k steps.

We prepared the following GAN-based vocoders for comparison: MelGAN, Parallel WaveGAN, and HiFi-GAN. We used the official implementations of MelGAN\footnote{https://github.com/descriptinc/melgan-neurips} and HiFi-GAN V1\footnote{https://github.com/jik876/hifi-gan} for reproducibility, and we implemented Parallel WaveGAN as per the process followed by Yamamoto et al\cite{yamamoto2020parallel}. The learning rate, optimizer, and all other parameters required for training followed the reference configurations of each model. As noted in section 4.1, all experiments were performed under the full-band conditions 
(0--12 kHz log-mel-spectrograms at 24 kHz sampling rate), unlike the band-limited conditions (0--8 kHz log-mel-spectrograms at 22.05 kHz sampling rate) used in the experimental setups in previous works\cite{kumar2019melgan, yamamoto2020parallel, kong2020hifi}.

All models in the experiments were trained to a mini-batch size of 32 using four NVIDIA V100 GPUs. To ensure stable sound quality, we trained up to 2M steps for MelGAN and up to 1M steps for the other models. 

For text-to-speech evaluation, we used the JDI-T\cite{lim2020jdi} acoustic model with a pitch and energy predictor\cite{ren2020fastspeech, lancucki2021fastpitch}. We converted text to phoneme sequences using open-sourced software\footnote{https://github.com/bootphon/phonemizer}. Each trained vocoder was fine-tuned up to 100k steps using ground truth waveforms and predicted log-mel-spectrograms.

\section{Results}

\subsection{Ablation study}
\begin{table}[t]
\centering
\caption{MOS of the ablation study.}
\renewcommand{\tabcolsep}{1.5mm}
\begin{tabular}{l|ccccc|c}
\toprule
Instance  & G1         & G2         & D1         & D2         & D3         & MOS                        \\
\midrule
Recordings & - & - & - & - & - &  4.16$\pm$0.09          \\
\midrule
UnivNet-c16 & \checkmark & \checkmark & \checkmark & \checkmark &            &  \textbf{3.92$\pm$0.08} \\
\midrule
Without LVC\tablefootnote{We used local conditioning\cite{van2016wavenet} instead of LVC and changed the channel sizes of the input layer and 3 residual stacks to 512, 256, 128, and 64, respectively.} &            & \checkmark & \checkmark & \checkmark &            &  3.73$\pm$0.10          \\
Without GAU & \checkmark &            & \checkmark & \checkmark &            &  3.82$\pm$0.09          \\
Without MRSD & \checkmark & \checkmark &            & \checkmark &            &  3.38$\pm$0.11 \\
Without MPWD & \checkmark & \checkmark & \checkmark &            &            &  3.15$\pm$0.11 \\
\midrule
MPWD$\to$MSWD & \checkmark & \checkmark & \checkmark &            & \checkmark & 3.58$\pm$0.10           \\
With MSWD & \checkmark & \checkmark & \checkmark & \checkmark & \checkmark & 3.90$\pm$0.09          \\
\bottomrule
\end{tabular}
\end{table}

\begin{figure}[t]
  \centering
  \includegraphics[width=0.96\linewidth]{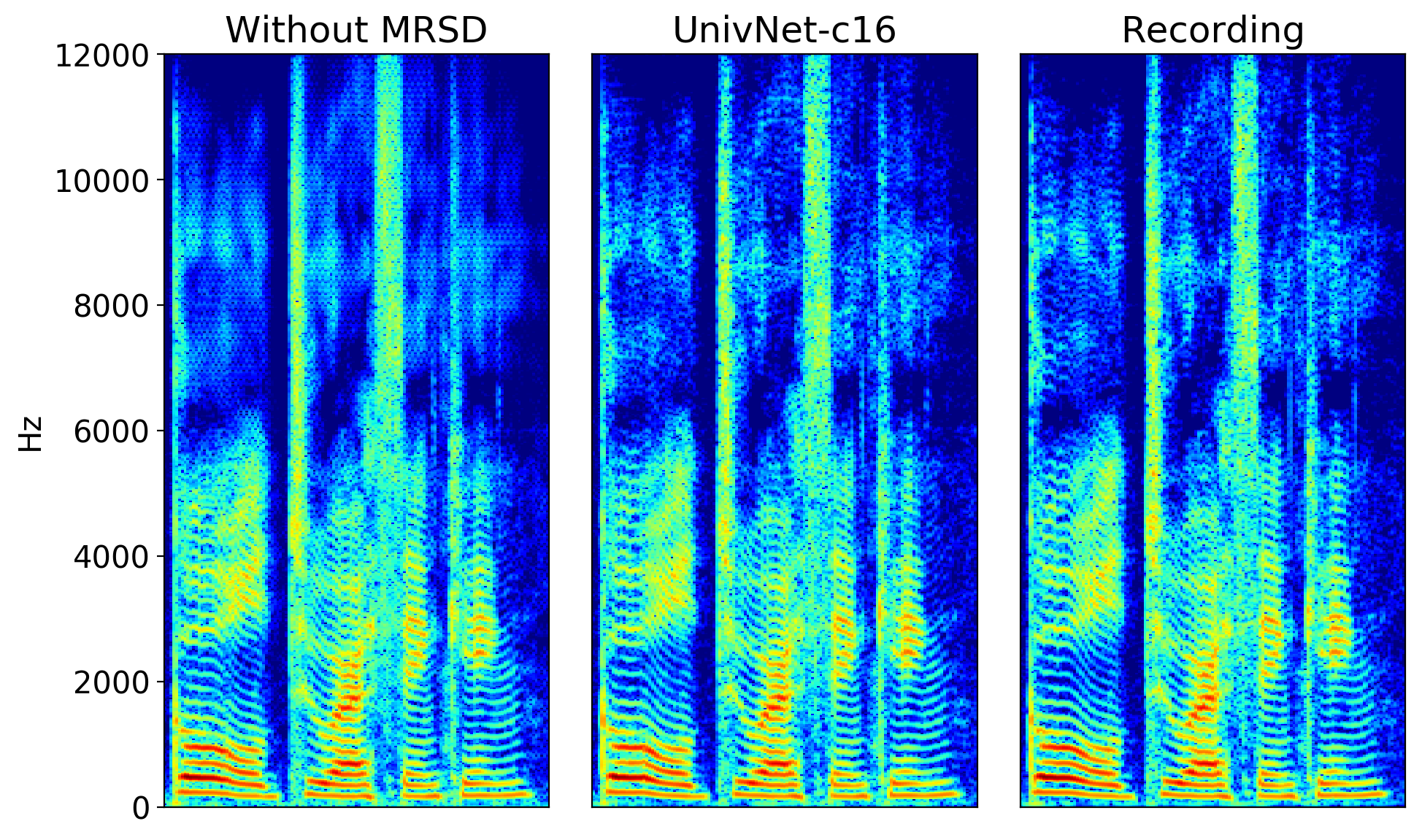}
  \caption{Spectrograms of the generated audio clips.}
\end{figure}

\begin{table*}[b]
\centering
\caption{Results of comparison with existing models. ‘\textup{Speed}’ indicates each model's generation speed relative to real time.}
\begin{tabular}{l|ccc|ccc|c|cc}
\toprule
                 & \multicolumn{3}{c|}{Seen speakers} & \multicolumn{3}{c|}{Unseen speakers} & Text-to-speech & \multicolumn{2}{c}{Model complexity} \\
\midrule
Model            &     MOS       & PESQ &  RMSE &      MOS      & PESQ &  RMSE &      MOS      & Params &          Speed \\
\midrule
MelGAN           & 3.56$\pm$0.10 & 2.80 & 0.576 & 2.89$\pm$0.11 & 2.74 & 0.546 & 2.66$\pm$0.12 &  4.34M & $\times$294.11 \\
Parallel WaveGAN & 3.07$\pm$0.10 & 3.60 & 0.357 & 2.61$\pm$0.11 & 3.61 & 0.332 & 2.85$\pm$0.11 &  1.44M &  $\times$66.23 \\
HiFi-GAN         & 3.89$\pm$0.09 & 3.54 & 0.423 & 3.86$\pm$0.09 & 3.55 & 0.389 & 3.94$\pm$0.08 & 14.01M & $\times$135.14 \\
UnivNet-c16      & 3.92$\pm$0.08 & 3.59 & 0.337 & 3.79$\pm$0.09 & 3.60 & 0.313 & 3.87$\pm$0.09 &  4.00M & $\times$227.27 \\
UnivNet-c32      & \textbf{3.93$\pm$0.09} & \textbf{3.70} & \textbf{0.316} & \textbf{3.90$\pm$0.09} & \textbf{3.70} & \textbf{0.294} & \textbf{4.07$\pm$0.08} & 14.86M & $\times$204.08 \\
\midrule
Recordings       & 4.16$\pm$0.09 & 4.50 & 0.000 & 4.30$\pm$0.08 & 4.50 & 0.000 &             - &      - &               - \\
\bottomrule
\end{tabular}
\end{table*}
To demonstrate the validity of the proposed model configuration, we prepared instances in which each component of the model was removed. We also prepared some combinations of discriminators, including MSWD\footnote{We used the settings employed by Kumar et al\cite{kumar2019melgan}.}. UnivNet-c16, a lightweight version of the model, was used for comparison.

In Table 1, where the MOS of the ablation study are listed, the components of the generator and discriminators are represented by the following index symbols for brevity: G1=LVC, G2=GAU, D1=MRSD, D2=MPWD, and D3=MSWD. A comparison of the above four instances and UnivNet-c16 shows that each component of the proposed model contributes to improving the performance. In particular, if MRSD is removed, an over-smoothing problem occurs, as shown in Figure 3. This is most notable in the high-frequency band of the generated waveform. The MOS of the instance is significantly reduced because of an audible metallic artifact. Further, the reduced score for the instance of removing MPWD indicates that the model supports the generator's detailed prediction in the temporal domain.

The comparison of the two instances below and UnivNet-c16 clarifies the reason why we chose MPWD instead of MSWD; the replacement of MPWD with MSWD significantly reduced the MOS. Despite the lower training speed, the instance of adding MSWD showed little difference from UnivNet-c16.

\subsection{Comparison with existing models}
We compared the sound quality and model complexity of the two versions of UnivNet with various vocoders. Each model's generation speed was measured on an NVIDIA V100 GPU.

The results of the comparison are summarized in Table 2. The first and second columns represent the sound quality of the seen and unseen speakers. MelGAN and Parallel WaveGAN achieved lower scores than the other models did, and their scores decreased significantly for the unseen speakers. Notably, the over-smoothing problem occurred in Parallel WaveGAN, because the experiments involving full-band features were performed, rather than those with band-limited features\cite{yamamoto2020parallel}. HiFi-GAN and UnivNet showed robustness by maintaining scores for both seen and unseen speakers. They demonstrated the possibility of generating high-quality waveforms for arbitrary speakers. This trend was also maintained in the text-to-speech scenario in the Text-to-speech column, which demonstrates that the model can rapidly adapt to a new speaker through fine-tuning.

The overall results demonstrate that UnivNet-c16 has a relatively small number of parameters and a speed that is 200 times higher than real time; further, UnivNet-c16 consistently maintains high objective and subjective scores. UnivNet-c32 achieves the highest objective scores in all scenarios, with only a slight decrease in speed. Compared with HiFi-GAN, the subjective scores show little difference, but UnivNet-c32 has the advantage that it can generate results approximately 1.5 times faster for a similar number of parameters.

\section{Conclusion}
In this paper, we proposed UnivNet, a vocoder with a multi-resolution spectrogram discriminator for high-fidelity and real-time waveform generation. The discriminator accepts multiple spectrograms as inputs and improves the spectral resolutions of the waveforms by alleviating the over-smoothing problem. In the future, we will research a universal vocoder for use in a multi-speaker text-to-speech pipeline without fine-tuning.

\bibliographystyle{IEEEtran}

\bibliography{mybib}

\end{document}